\documentclass{appolb}
\usepackage{epsfig}
\usepackage{cite}

\begin{document}
 \eqsec  
\title{ PHENOMENOLOGICAL STUDIES OF TOP PAIR \\ PRODUCTION AT
  NEXT-TO-LEADING ORDER \thanks{Presented at the XXXV International
    Conference of Theoretical Physics, Matter to the  Deepest: Recent
    Developments in Physics of Fundamental Interactions,  Ustron11, 12
    - 18 September 2011, Ustron, Poland}~\thanks{Preprint number: WUB/11-17}}%
\author{Ma\l{}gorzata Worek
\address{Fachbereich C Physik, Bergische Universit\"at Wuppertal
  \\ D-42097 Wuppertal,  Germany \\
\vspace{0.2cm}
 email: {\tt worek@physik.uni-wuppertal.de}}} 
\maketitle
\begin{abstract}
The calculation of NLO QCD corrections to the $t\bar{t}\to
W^{+}W^{-}b\bar{b}\to e^{+}\nu_e \mu^{-}\bar{\nu}_{\mu}b\bar{b}$
process with complete off-shell effects,  is briefly
summarized. Besides the total cross section and its scale dependence,
a few differential distributions  at the TeVatron run II and  LHC are
given. All results presented in this contribution have been obtained
with  the help of the \textsc{Helac-NLO} Monte Carlo framework.
\end{abstract}
  
\section{Introduction}

The Large Hadron Collider (LHC), with its two main multipurpose detectors
ATLAS and CMS, is the experimental project that dominates present
particle physics and will likely dominate its next 20-25 years.  With the
successful start of collisions at 7 TeV, the LHC has put yet another
big step towards  a thorough examination of the Terascale.
Ultimately, it has replaced the older, lower-energy TeVatron,
which has  been closed in September this year.  The large energy
available at the LHC has opened many multi-particle channels that
are now to a large degree scrutinized. The immense amount of
available phase space, and the large acceptance of the ATLAS and CMS
detectors allow for the production and identification of final states
with 4 or more QCD jets together with isolated leptons. These
multi-particle events hide or strongly modify all possible signals of
physics beyond the Standard Model.  In view of a correct
interpretation  of the signals of new physics which might be extracted
from data, it is of considerable interest to reduce our theoretical
uncertainty for the physical processes under study, especially when
large QCD backgrounds are involved. In this respect, the need of
next-to-leading-order (NLO) corrections for the LHC is
unquestionable. 

Efficient numerical evaluation of multi-particle final states at
NLO QCD can be performed with the help of the \textsc{Helac-NLO} Monte
Carlo program \cite{Bevilacqua:2011xh}. \textsc{Helac-NLO} is an
extension of  the \textsc{Helac-Phegas} Monte Carlo program
\cite{Kanaki:2000ey, Papadopoulos:2000tt,Cafarella:2007pc}, which is
based on off-shell Dyson-Schwinger recursive equations. It can be used
to efficiently obtain helicity amplitudes and total cross sections for
arbitrary multiparticle processes in the Standard Model and has been
already extensively used and tested, see {\it e.g.}
\cite{Gleisberg:2003bi,Papadopoulos:2005ky,Alwall:2007fs,
  Englert:2008tn,Actis:2010gg,Calame:2011zq}.   Virtual corrections
are obtained using the \textsc{Helac-1Loop} program
\cite{vanHameren:2009dr}, based on the Ossola-Papadopoulos-Pittau
(OPP) reduction technique \cite{Ossola:2006us} and the reduction code
\textsc{CutTools} \cite{Ossola:2007ax,Draggiotis:2009yb,
  Garzelli:2009is,Garzelli:2010qm,Garzelli:2010fq}. Moreover, the
\textsc{OneLOop}  library \cite{vanHameren:2010cp}  has been used for
the evaluation of the scalar integrals.  Reweighting techniques,
helicity and colour sampling methods  are used in order to optimize
the performance of the system. In addition, the singularities from
soft or collinear parton emission are isolated via Catani-Seymour 
dipole subtraction for NLO QCD calculations using a formulation for 
massive quarks  \cite{Catani:1996vz,Catani:2002hc}  and for arbitrary 
polarizations \cite{Czakon:2009ss}.  Calculations of this part  
are performed with the help of the \textsc{Helac-Dipoles} software
\cite{Czakon:2009ss}. The optimization and phase space integration is
executed with the help of \textsc{Parni} \cite{vanHameren:2007pt} and
\textsc{Kaleu} \cite{vanHameren:2010gg}. All parts of the \textsc{Helac-NLO}
framework are publictly available 
\footnote{\tt http://helac-phegas.web.cern.ch/helac-phegas/}.

\begin{figure}
\begin{center}
\includegraphics[width=0.8\textwidth]{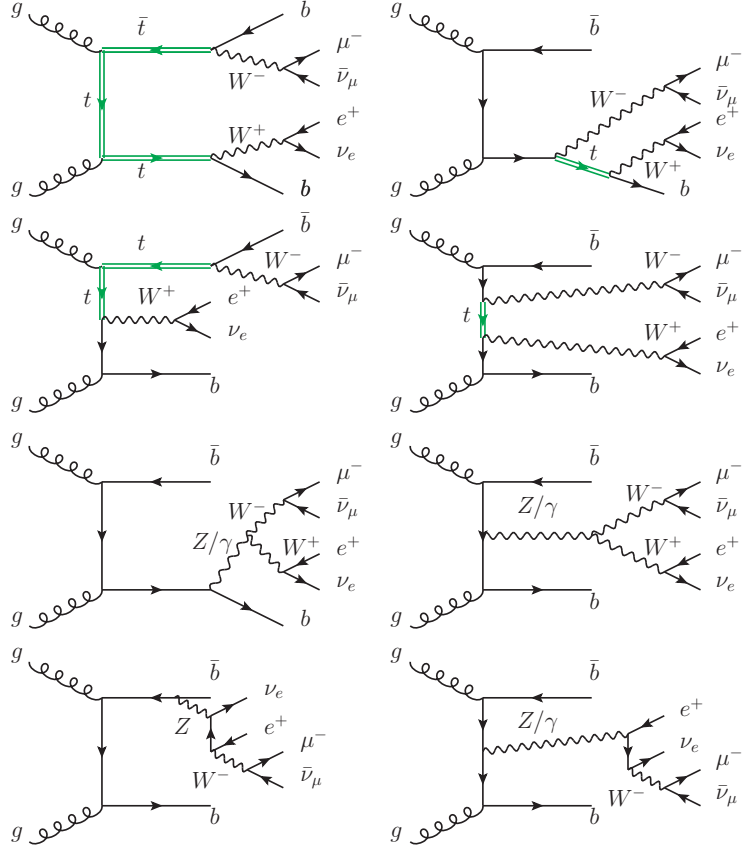}
\end{center}
\caption{\it \label{fig:diagrams-lo}   Representative Feynman diagrams
  contributing to the leading order process $gg\to
  e^{+}\nu_{e}\mu^{-}\bar{\nu}_{\mu}b\bar{b} $  at
  ${\cal{O}}(\alpha_{s}^2 \alpha^4)$,  with different off-shell
  intermediate states:  double-, single-, and non-resonant
  top quark contributions.}
\end{figure}

With the help of the \textsc{Helac-NLO} system several $2\to 4$ processes
have  recently been calculated at next-to-leading order QCD, including
$t\bar{t}b\bar{b}$ \cite{Bevilacqua:2009zn},  $t\bar{t}jj$
\cite{Bevilacqua:2010ve,Bevilacqua:2011hy}  and $W^{+}W^{-}b\bar{b}$
\cite{Bevilacqua:2010qb}.  In this contribution, a brief report on the
$pp(p\bar{p})\to t\bar{t}\to W^{+}W^{-}b\bar{b}\to
e^{+}\nu_{e}\mu^{-}\bar{\nu}_{\mu}b\bar{b}$  computation with complete
off-shell effects is given. Double-, single- and non-resonant top
contributions of  the order ${\cal{O}}(\alpha_{s}^3 \alpha^4)$  are
consistently taken into account, which requires the introduction of a
complex-mass scheme for unstable  top quarks.  Moreover, the
intermediate $W$ bosons are treated off-shell. A few examples of
Feynman diagrams  contributing to the leading order $gg\to
e^{+}\nu_{e}\mu^{-}\bar{\nu}_{\mu}b\bar{b}$ subprocess are presented
in Figure \ref{fig:diagrams-lo}. 

Parallel to our work, another NLO study of $t\bar{t}b\bar{b}$
\cite{Bredenstein:2008zb,Bredenstein:2009aj,Bredenstein:2010rs}  at 
the  LHC appeared. Moreover, NLO QCD corrections to the   
$W^{+}W^{-}b\bar{b}$ \cite{Denner:2010jp} process have been calculated.

\section{Numerical Results}

The process $pp(p\bar{p}) \rightarrow  t\bar{t} + X
\rightarrow  W^+W^-b\bar{b} + X \rightarrow e^+ \nu_e \mu^- \nu_{\mu}
b\bar{b} +X$  is considered, 
both at the TeVatron run II and the LHC {\it i.e.}  at a
center-of-mass energy of $\sqrt{s} = 1.96$ TeV and $\sqrt{s} = 7$ TeV
correspondingly. The
Standard Model parameters are  as follows:
\begin{equation}
m_W = 80.398 ~\textnormal{GeV}, ~~~~ \Gamma_{W}=2.141 ~\textnormal{GeV}
\end{equation}
\begin{equation}
m_Z=91.1876  ~\textnormal{GeV}, ~~~~ \Gamma_Z=2.4952 ~\textnormal{GeV}
\end{equation}
\begin{equation}
G_\mu = 1.16639 \times 10^{-5} ~\textnormal{GeV}^{-2}
\end{equation}
The electromagnetic coupling and $\sin^2\theta_W$ are derived from the
Fermi  constant and masses of W and Z bosons. The top quark mass is
$m_t = 172.6 $ GeV and  all other QCD partons and leptons
are treated as massless. The top quark width is $\Gamma_{t}^{LO}=1.48
~\mbox{GeV}$ at LO and $\Gamma_{t}^{NLO}=1.35 ~\mbox{GeV}$ at NLO
where $\alpha_s = \alpha_s(m_{t})= 0.107639510785815$.  The on-shell
scheme is adopted for mass renormalization.  All final-state
partons  with pseudorapidity $|\eta| < 5$ are  recombined into jets
via  the $k_T$ algorithm \cite{Catani:1992zp,Catani:1993hr,Ellis:1993tq}, 
the {\it anti-}$k_T$ algorithm \cite{Cacciari:2008gp} 
and the inclusive Cambridge/Aachen algorithm (C/A) \cite{Dokshitzer:1997in} 
with a cone of size 
$R  = 0.4$. Additional cuts are imposed on the transverse momenta and
the rapidity of two recombined $b$-jets: 
\begin{equation}
p_{T_{b}} > 20 ~\textnormal{GeV},  ~~~~~~|y_b|< 4.5.
\end{equation}  
Basic selection  is applied to decay products of top quarks:
\begin{equation}
p_{T_{\ell}}> 20 ~\textnormal{GeV}, ~~|\eta_\ell| < 2.5, ~~\Delta
R_{b\ell} > 0.4,  ~~p_{T_{miss}} > 30 ~\textnormal{GeV}.
\end{equation}  
The
CTEQ6 set of parton  distribution functions (PDFs) is consistently used 
\cite{Pumplin:2002vw,Stump:2003yu}. 
In particular,   CTEQ6L1 PDFs with a 1-loop running $\alpha_s$ is taken
at LO and  CTEQ6M PDFs with a 2-loop running $\alpha_s$ at NLO. The
contribution  from $b$ quarks in the initial state is neglected.  The
number of active flavors is $N_F = 5$, and the respective QCD
parameters  are $\Lambda^{LO}_5 = 165$ MeV and  $\Lambda^{MS}_5 = 226$
MeV. In the renormalization of the strong coupling constant, the
top-quark loop in the gluon self-energy is subtracted at zero
momentum. In this scheme the running of $\alpha_s$ is generated by the
contributions of the light-quark and gluon loops. The renormalization
and factorization scales, $\mu_{R}$  and $\mu_F$, are set to the
common value $\mu =m_t$. 
\begin{table}[t!]
\begin{center}
  \begin{tabular}{l|c|r}
      Algorithm & $\sigma_{\rm LO}$ [fb]      &   
      $\sigma_{\rm NLO}$   [fb] \\
&&\\ \hline &&\\
{\it anti}-$k_T$ 
& 34.922 $\pm$ 0.014 & 35.697 $\pm$ 0.049 \\
$k_T$ &  34.922 $\pm$ 0.014 & 35.723 $\pm$  0.049   \\
C/A &  34.922 $\pm$ 0.014 & 35.746  $\pm$ 0.050    \\
  \end{tabular}
\end{center}
  \caption{\it \label{tab:tev} Integrated cross section at LO and NLO
    for  $p\bar{p}\rightarrow
    e^{+}\nu_{e}\mu^{-}\bar{\nu}_{\mu}b\bar{b} ~+ X$  production at
    the TeVatron run II. }
\end{table}
\begin{figure}
\begin{center}
\includegraphics[width=0.49\textwidth]{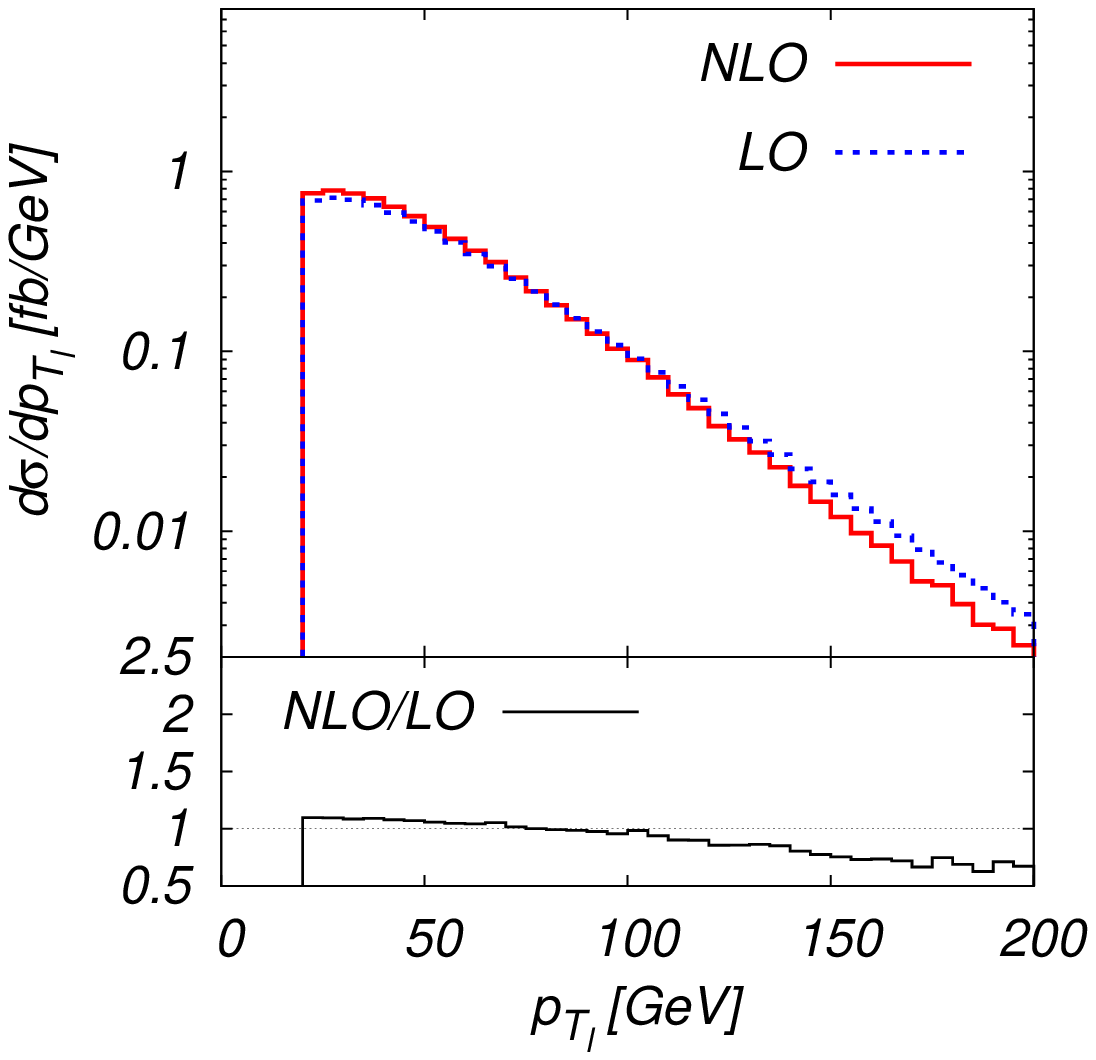}
\includegraphics[width=0.49\textwidth]{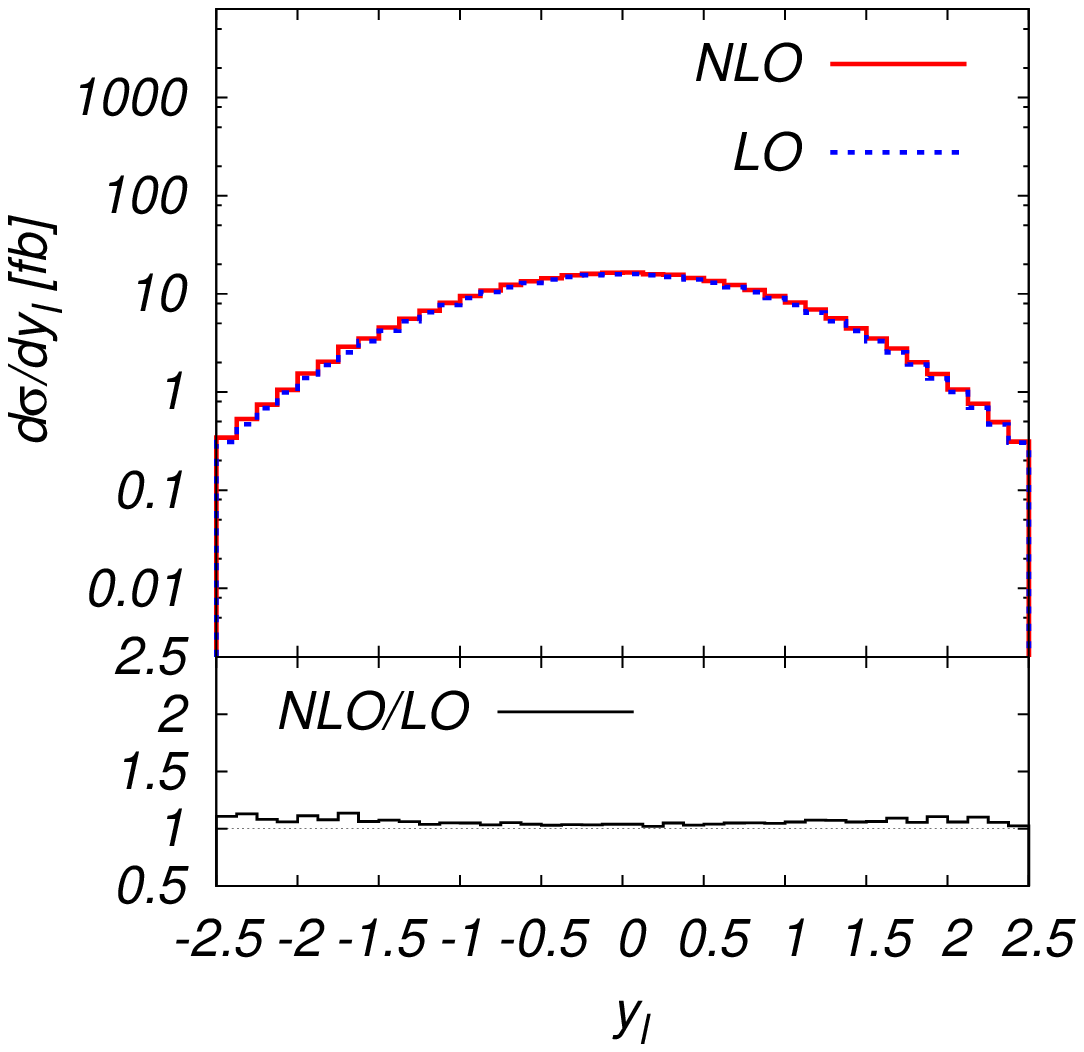}
\includegraphics[width=0.49\textwidth]{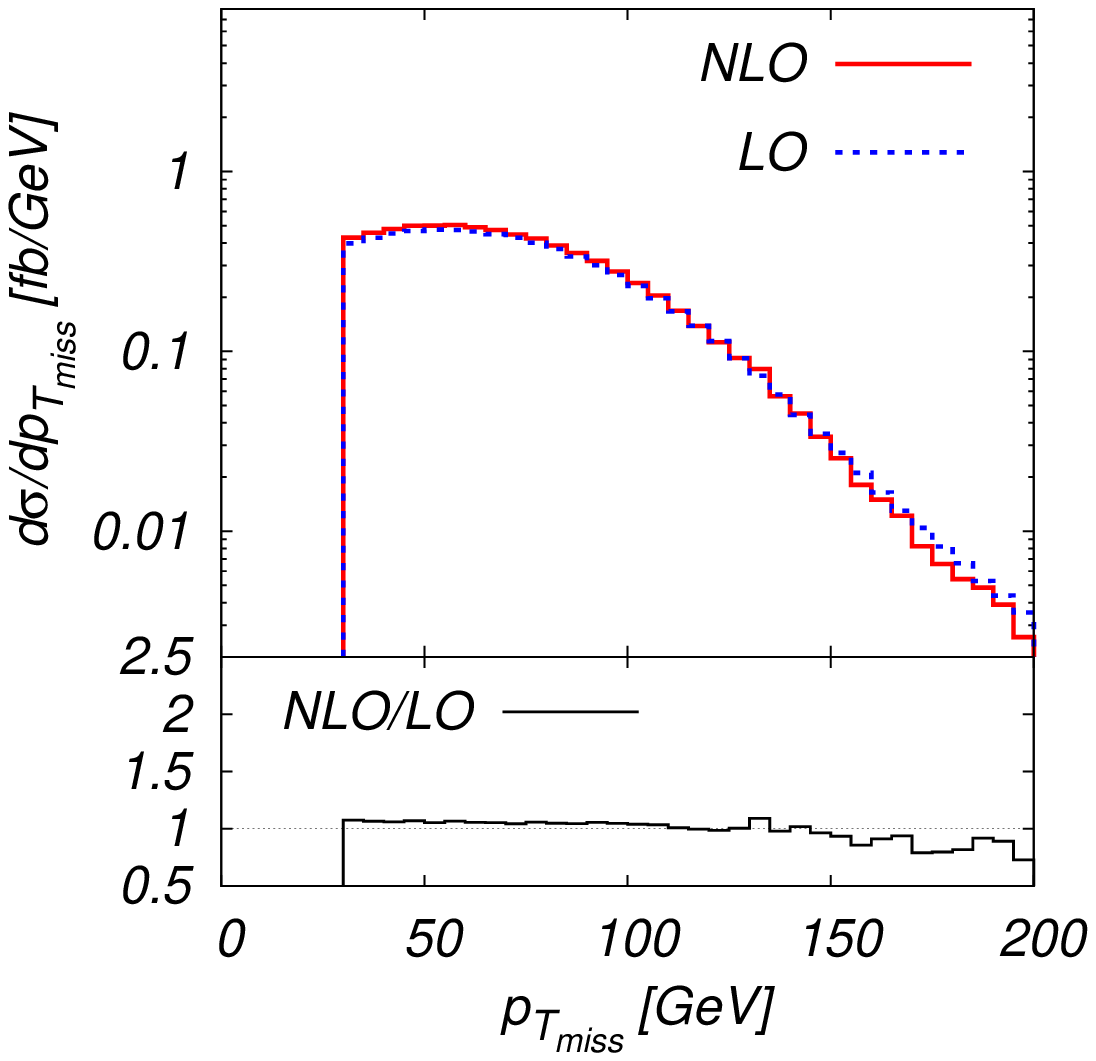}
\includegraphics[width=0.49\textwidth]{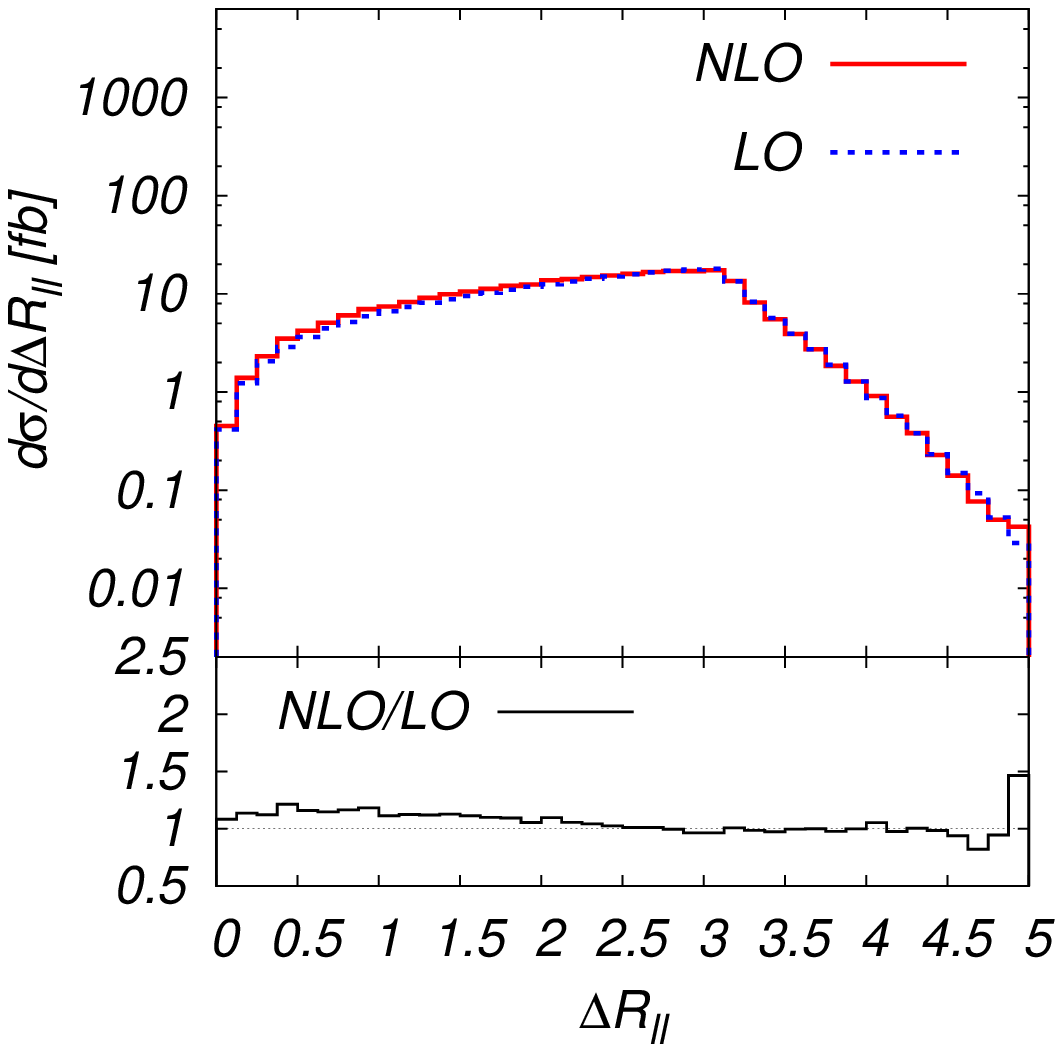}
\end{center}
\caption{\it \label{fig:lepton-tev}  Differential  cross section
  distributions as a function of the averaged transverse momentum
  $p_{T_{\ell}}$  of the  charged leptons,  averaged  rapidity
  $y_{\ell}$ of the  charged leptons, $p_{T_{miss}}$ and $\Delta
  R_{\ell\ell}$  for the $p\bar{p}\rightarrow
  e^{+}\nu_{e}\mu^{-}\bar{\nu}_{\mu}b\bar{b} ~ + X$ process at the
  TeVatron run II.  The blue dashed curve corresponds to the leading
  order, whereas the red solid one to the next-to-leading order
  result. The lower panels display  the differential K factor. }
\end{figure}

\subsection{TeVatron Run II}

We start 
with a discussion of the total cross section at the TeVatron run II.
In spite of the fact that the TeVatron 
has been recently closed, the data analysis in the
 CDF and D0 experiments is still ongoing. Therefore, in Table
\ref{tab:tev} results for the total cross section for the
central  value of the scale, $\mu_R = \mu_F = m_t$ and for three
different jet  algorithms:  $k_T$,  {\it anti-}$k_T$ and the
inclusive Cambridge/Aachen algorithm (C/A), are presented. 
The total cross  section receives small NLO QCD
correction of the order of 2\%.  Residual scale uncertainties, as obtained
by varying the scale down and up by a
factor 2, are at the 40\% level in the LO case.
The dependence is large, illustrating the well known fact that  the LO
prediction can only provide a rough estimate.  As expected, we observe
a reduction of the scale uncertainty while going from LO to NLO. In
the NLO case we have obtained a variation of the order of 8\%.
In adition, the size of the non-factorizable correctionsm, as obtained by a
comparison of the full result against a result in the narrow width
approximation (NWA), amounts to 1\%. This is consistent with the uncertainty
of the NWA {\it i.e.} which is of order  ${\cal {O}}(\Gamma_t/m_t)$.

In the next step, corrections to differential distributions are
presented.  In Figure \ref{fig:lepton-tev}, differential  cross
section distributions as function of the averaged transverse momentum
and  averaged rapidity of the  charged leptons are given. Also shown are
distributions of missing transverse momentum,  $p_{T_{miss}}$, and 
dilepton separation in the azimuthal angle-pseudorapidity plane, 
$\Delta R_{\ell\ell}$.  Even though
the NLO corrections to the transverse   momentum distribution are
moderate, they do not simply rescale the   LO shape. A distortion at
the level of $40\%$ is induced. For $p_{T_{miss}}$, a distortion only up
to  $15\%$ can be observed.  As for angular  distributions positive
and rather modest corrections of the order  of $5\%-10\%$ are
obtained.

\subsection{Large Hadron Collider}

\begin{table}
\begin{center}
  \begin{tabular}{l|c|r}
     Algorithm & $\sigma_{\rm LO}$ [fb]      &
    $\sigma_{\rm NLO}$   [fb] \\ 
    & & \\ \hline
   & & \\
   {\it anti}-$k_T$  & 550.54 $\pm$ 0.18 & 
    808.29 $\pm$ 1.04 \\  $k_T$ & 550.54 $\pm$ 0.18 
   & 808.86 $\pm$ 1.03 \\  C/A & 550.54 $\pm$
    0.18  & 808.28 $\pm$ 1.03  
  \end{tabular}
\end{center}
 \caption{\it \label{tab:lhc} Integrated cross section at LO and NLO
   for  $pp\rightarrow e^{+}\nu_{e}\mu^{-}\bar{\nu}_{\mu}b\bar{b} ~+
   X$  production at the LHC.}
\end{table}
\begin{figure}
\begin{center}
\includegraphics[width=0.49\textwidth]{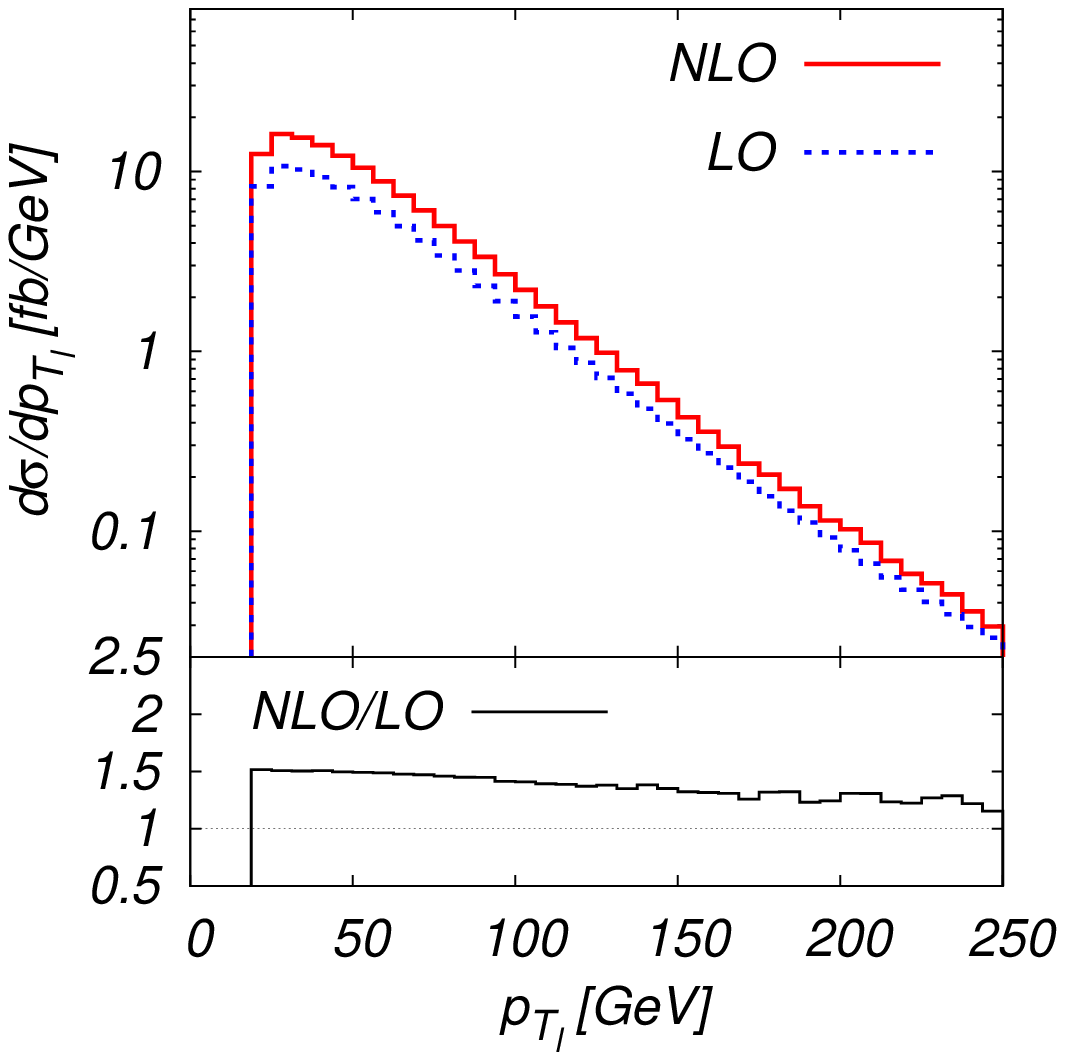}
\includegraphics[width=0.49\textwidth]{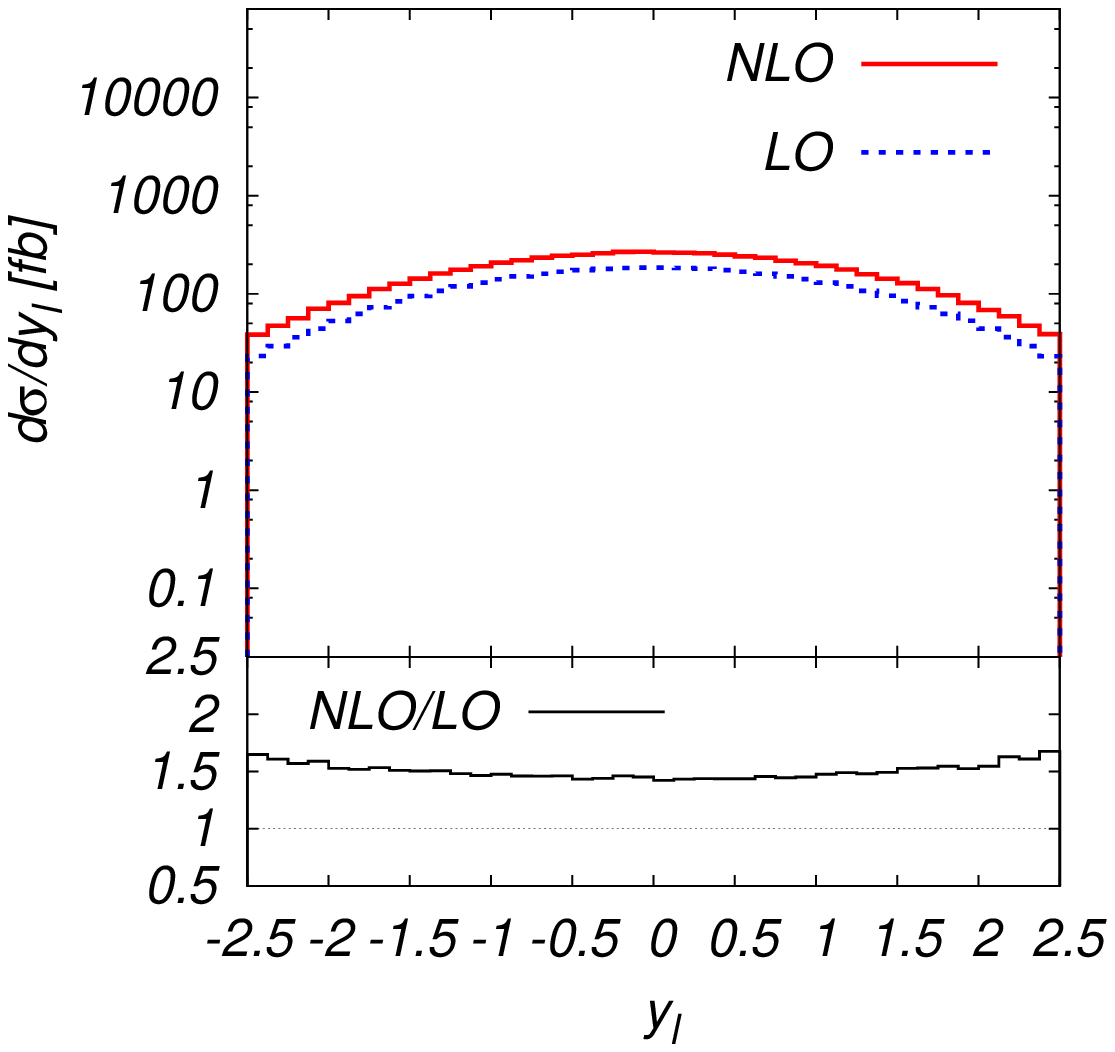}
\includegraphics[width=0.49\textwidth]{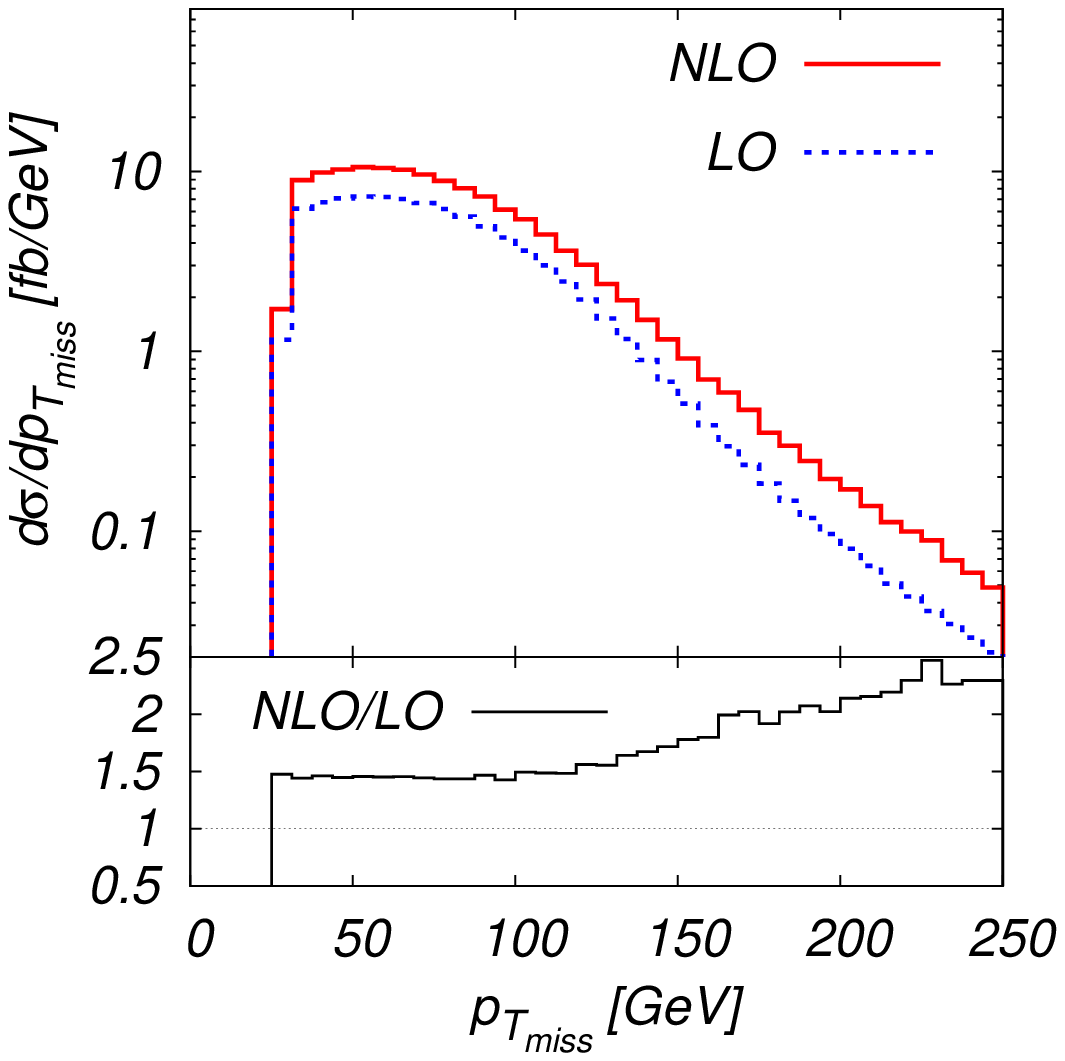}
\includegraphics[width=0.49\textwidth]{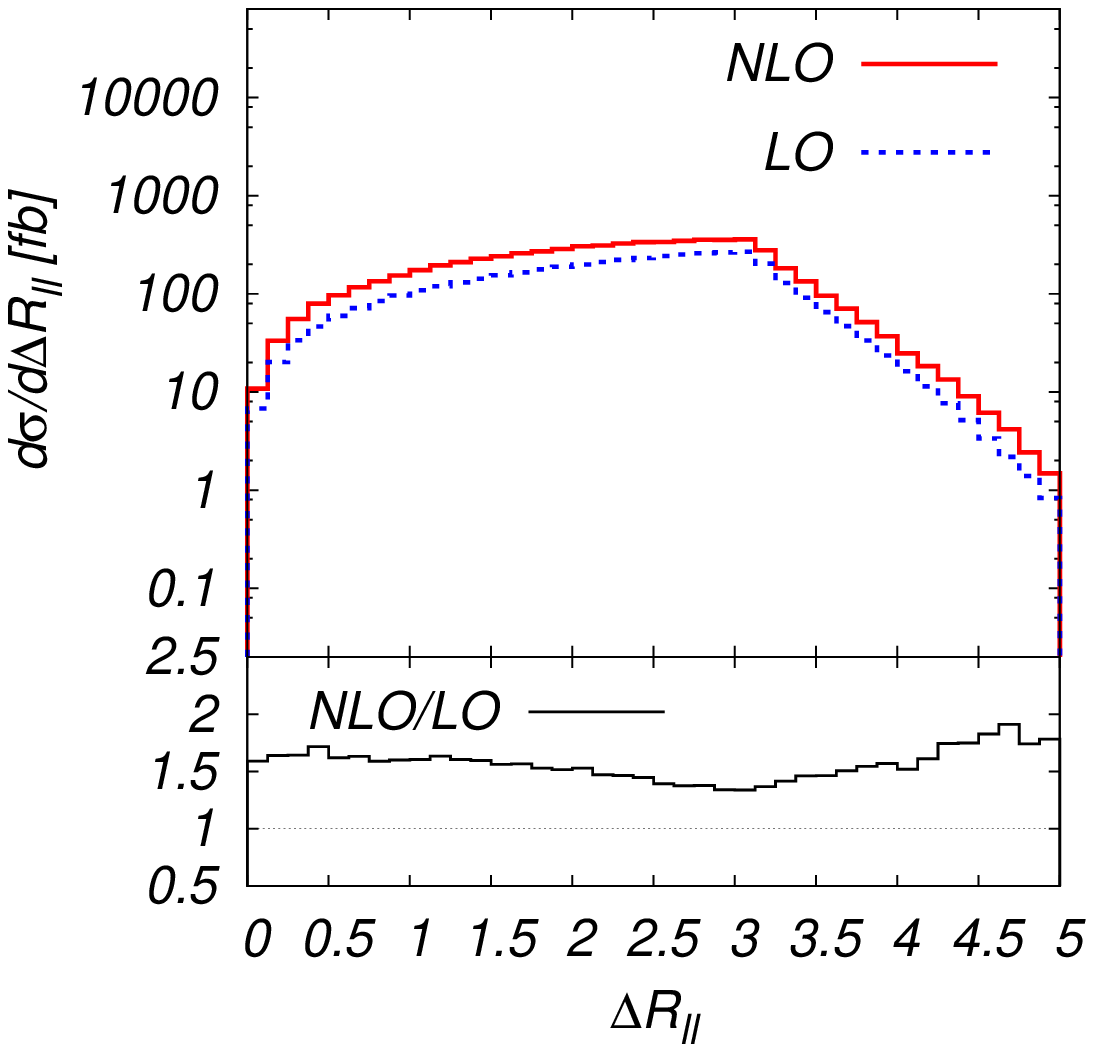}
\end{center}
\caption{\it \label{fig:leptons-lhc} Differential  cross section
  distributions as a function of  the averaged transverse momentum
  $p_{T_{\ell}}$  of the  charged leptons,  averaged  rapidity
  $y_{\ell}$ of the  charged leptons, $p_{T_{miss}}$  and $\Delta
  R_{\ell\ell}$  for the $pp\rightarrow
  e^{+}\nu_{e}\mu^{-}\bar{\nu}_{\mu}b\bar{b} ~ + X$ process at the LHC. 
  The blue dashed curve corresponds to the
  leading order, whereas the red solid one to the next-to-leading
  order result. The lower panels display  the differential K factor. }
\end{figure}

Table \ref{tab:lhc} shows the integrated cross sections at the LHC
with $\sqrt{s}= 7$ TeV, for three different jet algorithms.  At the
central scale value, the full cross section receives  NLO QCD
corrections  of the order of $47\%$. After including the NLO corrections, a
large scale dependence of about $37\%$ in the LO cross section is considerably
reduced down to $9\%$.

In order to quantify the size of the non-factorizable corrections for
the LHC, a comparison to the narrow-width limit of our
calculation has again been performed. Going
from NWA to the full result changes the cross section no more than
$1.2\%$ for our inclusive setup.

in Figure \ref{fig:leptons-lhc}, differential  cross section
distributions as function of the averaged transverse momentum and
averaged  rapidity of the  charged leptons together with $p_{T_{miss}}$ and
$\Delta R_{\ell\ell}$ separation  are shown. 

For renormalization and factorization scales set to a common value
$\mu=m_t$, the NLO QCD corrections are  always positive and relatively
large.  In particular, in case of the $p_{T_{\ell}}$ differential
distribution, a distortion up to $25\%$ is reached, while for
$p_{T_{miss}}$ a distortion up to  $80\%$ is visible.  For the
$y_{\ell}$ distribution,  rather constant corrections up to $50\%$
are obtained. And finally, the distribution in $\Delta R_{\ell\ell}$
has even acquired  corrections up to $90\%$.

\section{Summary}

The NLO QCD corrections to the full  decay  chain
$pp(p\bar{p}) \rightarrow t \bar{t}\to W^+W^- b\bar{b} \to
e^{+}\nu_{e} \mu^{-}\bar{\nu}_{\mu}b\bar{b} ~+X$ have been briefly
presented.  In the calculation, all  off-shell effects of top quarks
and $W$  gauge  bosons have been included in a fully differential
way. The total cross section and its scale dependence, as well as a few
differential distributions at the TeVatron run II and the LHC have
been given.  The  impact of the NLO QCD corrections  on integrated
cross sections at the TeVatron  is small,  of the order $2 \%$. On the 
other hand, at the LHC, $47\%$ NLO QCD corrections have been obtained.
Residual theoretical uncertainties due to higher order corrections have
 been estimated to be  at the $8\%-9\%$ level. An finally,
NLO QCD corrections do not only affect the
overall normalization of the integrated cross sections, but can 
also change the shape  of some differential distributions. 

\section*{Acknowledgment}

I would like to thank the organizers of the XXXV International
Conference of Theoretical Physics, Ustron11,  for the kind invitation
and the very pleasant atmosphere during the conference.

The calculations presented here, have been performed on the Grid
Cluster of the Bergische Universit ̈at Wuppertal, financed by the
Helmholtz - Alliance “Physics at the Terascale” and the BMBF.

The author was supported by the Initiative and Networking Fund of the
Helmholtz Association, contract HA-101 (“Physics at the Terascale”).



\begin{thebibliography}{}

\bibitem{Bevilacqua:2011xh}
G. Bevilacqua, M. Czakon, M.V. Garzelli, A. van Hameren, A. Kardos, C.G. 
Papadopoulos, R. Pittau, M. Worek, 
{\it HELAC-NLO}, 
{\tt  [arXiv:1110.1499 [hep-ph]]}.

\bibitem{Kanaki:2000ey}
A. Kanaki and C. G. Papadopoulos, 
{\it HELAC: A package to compute electroweak  helicity amplitudes},  
Comput. Phys. Commun. {\bf 132} (2000) 306, {\tt [hep-ph/0002082]}.

\bibitem{Papadopoulos:2000tt}
C. G. Papadopoulos, 
{\it PHEGAS: A phase space generator for automatic cross section computation},
Comput. Phys. Commun. {\bf 137} (2001) 247, {\tt [hep-ph/0007335]}.

\bibitem{Cafarella:2007pc}
A. Cafarella, C. G. Papadopoulos and M. Worek, 
{\it {Helac-Phegas: a generator for all parton level processes}},  
Comput. Phys. Commun. {\bf 180}  (2009) 1941, {\tt [arXiv:0710.2427 [hep-ph]]}.

\bibitem{Gleisberg:2003bi}
T. Gleisberg, F. Krauss, C. G. Papadopoulos, A. Schaelicke and S. Schumann,
{\it {Cross sections for multi-particle final states at a linear collider}},
Eur. Phys. J. {\bf C34} (2004) 173, {\tt [hep-ph/0311273]}.

\bibitem{Papadopoulos:2005ky}
C. G. Papadopoulos and M. Worek, 
{\it {Multi-parton Cross Sections at Hadron Colliders}},  
Eur. Phys. J. {\bf C50} (2007) 843, {\tt [hep-ph/0512150]}.

\bibitem{Alwall:2007fs}
J. Alwall {\em et. al.}, 
{\it {Comparative study of various algorithms for the  merging of parton 
showers and matrix elements in hadronic collisions}},  
Eur. Phys. J. {\bf C53} (2008) 473, {\tt [arXiv:0706.2569  [hep-ph]]}.

\bibitem{Englert:2008tn}
C. Englert, B. Jager, M. Worek and D. Zeppenfeld, 
{\it {Observing Strongly Interacting Vector Boson Systems at the CERN Large 
Hadron Collider}},  
Phys. Rev. {\bf D80} (2009) 035027, {\tt [arXiv:0810.4861  [hep-ph]]}.

\bibitem{Actis:2010gg}
S. Actis {\em et. al.}, 
{\it {Quest for precision in hadronic cross sections at low energy: 
Monte Carlo tools vs. experimental data}},  
Eur. Phys. J.  {\bf C66} (2010) 585, {\tt  [arXiv:0912.0749  [hep-ph]]}.

\bibitem{Calame:2011zq}
C. C. Calame {\it et al.},
{\it NNLO leptonic and hadronic corrections to Bhabha scattering and luminosity
monitoring at meson factories}, 
JHEP {\bf 1107}, 126 (2011), {\tt [arXiv:1106.3178 [hep-ph]]}.

\bibitem{vanHameren:2009dr}
A. van Hameren, C. G. Papadopoulos and R. Pittau, 
{\it {Automated one-loop calculations: a proof of concept}},  
JHEP {\bf 0909} (2009) 106, {\tt [arXiv:0903.4665  [hep-ph]]}.

\bibitem{Ossola:2006us}
G. Ossola, C. G. Papadopoulos and R. Pittau, 
{\it {Reducing full one-loop amplitudes to scalar integrals at the integrand 
level}},  
Nucl. Phys.  {\bf B763} (2007) 147, {\tt  [hep-ph/0609007]}.

\bibitem{Ossola:2007ax}
G. Ossola, C. G. Papadopoulos and R. Pittau, 
{\it {CutTools: a program implementing the OPP reduction method to compute 
one-loop amplitudes}},  
JHEP {\bf 0803} (2008) 042, {\tt  [arXiv:0711.3596  [hep-ph]]}.

\bibitem{Draggiotis:2009yb}
P. Draggiotis, M. V. Garzelli, C. G. Papadopoulos and R. Pittau, 
{\it {Feynman  Rules for the Rational Part of the QCD 1-loop amplitudes}},  
JHEP {\bf  0904} (2009) 072, {\tt [arXiv:0903.0356  [hep-ph]]}.

\bibitem{Garzelli:2009is}
M. V. Garzelli, I. Malamos and R. Pittau, 
{\it Feynman rules for the rational part of the Electroweak 1-loop amplitudes}, 
JHEP {\bf 1001} (2010) 040, [Erratum-ibid.\  {\bf 1010} (2010) 097],
{\tt   [arXiv:0910.3130 [hep-ph]]}.

\bibitem{Garzelli:2010qm}
M. V. Garzelli, I. Malamos and R. Pittau, 
{\it Feynman rules for the rational part of the Electroweak 1-loop 
amplitudes in the $R_\xi$ gauge and in the Unitary gauge},   
JHEP {\bf 1101} (2011) 029, {\tt [arXiv:1009.4302 [hep-ph]]}.

\bibitem{Garzelli:2010fq}
M. V. Garzelli and I. Malamos, 
{\it R2SM: A Package for the analytic computation of the $R_2$ Rational 
terms in the Standard Model of the Electroweak interactions},  
Eur. Phys. J. {\bf C71} (2011) 1605,
{\tt  [arXiv:1010.1248 [hep-ph]]}.

\bibitem{vanHameren:2010cp}
A.van Hameren, 
{\it OneLOop: For the evaluation of one-loop scalar functions},  
Comput. Phys. Commun.  {\bf 182} (2011) 2427, {\tt [arXiv:1007.4716 [hep-ph]]}.

\bibitem{Catani:1996vz}
S. Catani and M. H. Seymour, 
{\it {A general algorithm for calculating jet cross sections in NLO QCD}},  
Nucl. Phys. {\bf B485} (1997) 291, {\tt [hep-ph/9605323]}.

\bibitem{Catani:2002hc}
S. Catani, S. Dittmaier, M. H. Seymour and Z. Trocsanyi,  
{\it {The dipole formalism for next-to-leading order QCD calculations with 
massive partons}},
Nucl. Phys. {\bf B627} (2002) 189,
{\tt [hep-ph/0201036]}.

\bibitem{Czakon:2009ss}
M. Czakon, C. G. Papadopoulos and M. Worek, 
{\it {Polarizing the Dipoles}},
JHEP {\bf 0908} (2009) 085, {\tt  [arXiv:0905.0883  [hep-ph]]}.

\bibitem{vanHameren:2007pt}
A. van Hameren, 
{\it {PARNI for importance sampling and density estimation}},
Acta Phys. Polon. {\bf B40} (2009) 259, {\tt [arXiv:0710.2448  [hep-ph]]}.

\bibitem{vanHameren:2010gg}
A. van Hameren, 
{\it {Kaleu: a general-purpose parton-level phase space  generator}}, 
{\tt [arXiv:1003.4953  [hep-ph]]}.

\bibitem{Bevilacqua:2009zn}
G. Bevilacqua, M. Czakon, C. G. Papadopoulos, R. Pittau and M. Worek, {\it
{Assault on the NLO Wishlist: $pp \to t\bar{t} b\bar{b}$}},  JHEP {\bf
0909} (2009) 109, {\tt [arXiv:0907.4723  [hep-ph]]}.

\bibitem{Bevilacqua:2010ve}
G. Bevilacqua, M. Czakon, C. G. Papadopoulos and M. Worek, {\it {Dominant QCD
  Backgrounds in Higgs Boson Analyses at the LHC: A Study of
$pp \to t \bar{t}$  $+$ 2 jets at Next-To-Leading Order}},
Phys. Rev. Lett. {\bf 104}  (2010) 162002,
{\tt [arXiv:1002.4009  [hep-ph]]}.

\bibitem{Bevilacqua:2011hy}
G. Bevilacqua, M. Czakon, C. G. Papadopoulos and M. Worek, 
{\it Hadronic top-quark pair production in association with two jets at
Next-to-Leading Order QCD},  {\tt [arXiv:1108.2851 [hep-ph]]}.

\bibitem{Bevilacqua:2010qb}
G.~Bevilacqua, M.~Czakon, A.~van Hameren, C.~G.~Papadopoulos and M.~Worek,
{\it Complete off-shell effects in top quark pair hadroproduction with 
leptonic decay at next-to-leading order},  JHEP {\bf 1102} (2011) 083,
{\tt [arXiv:1012.4230 [hep-ph]]}.

\bibitem{Bredenstein:2008zb}
A. Bredenstein, A. Denner, S. Dittmaier and S. Pozzorini, 
{\it NLO QCD corrections to t anti-t b anti-b production at the LHC: 1.
Quark-antiquark annihilation},  JHEP {\bf 0808} (2008) 108, 
{\tt [arXiv:0807.1248 [hep-ph]]}.

\bibitem{Bredenstein:2009aj}
A.~Bredenstein, A.~Denner, S.~Dittmaier and S.~Pozzorini, 
{\it NLO QCD corrections to $pp \to t\bar{t}b\bar{b} + X$  at the LHC},
Phys.\ Rev.\ Lett.\  {\bf 103} (2009) 012002, 
{\tt [arXiv:0905.0110 [hep-ph]]}.

\bibitem{Bredenstein:2010rs}
A.~Bredenstein, A.~Denner, S.~Dittmaier and S.~Pozzorini, 
{\it NLO QCD corrections to top anti-top bottom anti-bottom production at the
LHC: 2. full hadronic results},  JHEP {\bf 1003} (2010) 021, 
{\tt [arXiv:1001.4006 [hep-ph]]}.

\bibitem{Denner:2010jp}
A.~Denner, S.~Dittmaier, S.~Kallweit and S.~Pozzorini, 
{\it NLO QCD corrections to WWbb production at hadron colliders},  
Phys.\ Rev.\ Lett.\  {\bf 106} (2011) 052001, 
{\tt [arXiv:1012.3975 [hep-ph]]}.

\bibitem{Catani:1992zp}
S. Catani, Y. L. Dokshitzer and B. R. Webber, {\it {The k-perpendicular
clustering algorithm for jets in deep inelastic scattering and hadron
collisions}},  Phys. Lett. {\bf B285} (1992) 291.

\bibitem{Catani:1993hr}
S. Catani, Y. L. Dokshitzer, M. H. Seymour and B. R. Webber, {\it
{Longitudinally invariant $k_T$ clustering algorithms for hadron hadron
collisions}},  Nucl. Phys. {\bf B406} (1993) 187.

\bibitem{Ellis:1993tq}
S. D. Ellis and D. E. Soper, {\it {Successive combination jet algorithm for
hadron collisions}},  Phys. Rev. {\bf D48} (1993) 3160,
{\tt [hep-ph/9305266]}.

\bibitem{Cacciari:2008gp}
M. Cacciari, G. P. Salam and G. Soyez, {\it {The anti-$k_T$ jet clustering
algorithm}},  JHEP {\bf 0804} (2008) 063,
{\tt  [arXiv:0802.1189  [hep-ph]]}.

\bibitem{Dokshitzer:1997in}
Y. L. Dokshitzer, G. D. Leder, S. Moretti and B. R. Webber, {\it {Better Jet
Clustering Algorithms}},  JHEP {\bf 9708} (1997) 001,
{\tt [hep-ph/9707323]}.

\bibitem{Pumplin:2002vw}
J. Pumplin {\em et. al.}, {\it {New generation of parton distributions with
uncertainties from global QCD analysis}},  JHEP {\bf 0207} (2002) 012,
{\tt [hep-ph/0201195]}.

\bibitem{Stump:2003yu}
D. Stump {\em et. al.}, {\it {Inclusive jet production, parton distributions,
and the search for new physics}},  JHEP {\bf 0310} (2003) 046,
{\tt [hep-ph/0303013]}.

\end{thebibliography}
\end{document}